# Model Checking of Statechart Models
## Survey and Research Directions

Purandar Bhaduri
TRDDC, Tata Consultancy Services
54 B, Hadapsar Industrial Estate
Pune 411 013, INDIA.
Email: `pbhaduri@pune.tcs.co.in`

S. Ramesh
Dept. of Computer Science & Engineering
Indian Institute of Technology Bombay
Powai, Mumbai 400 076, INDIA.
Email: `ramesh@cse.iitb.ac.in`

February 1, 2008

**Abstract**

We survey existing approaches to the formal verification of statecharts using model checking. Although the semantics and subset of statecharts used in each approach varies considerably, along with the model checkers and their specification languages, most approaches rely on translating the hierarchical structure into the flat representation of the input language of the model checker. This makes model checking difficult to scale to industrial models, as the state space grows exponentially with flattening. We look at current approaches to model checking hierarchical structures and find that their semantics is significantly different from statecharts. We propose to address the problem of state space explosion using a combination of techniques, which are proposed as directions for further research.

## 1 Introduction

Model checking [16] is a formal verification technique based on exhaustive state space exploration of the model of a finite state system (FSM). Given an input FSM model and a property in temporal logic, a model checker determines whether the property holds in the model, and returns with a counterexample trace in case the property fails.

In this paper we review various approaches to model checking of statecharts [21], which extend conventional state machines with hierarchy, concurrency and communication. We compare existing approaches to model checking of statecharts in the literature, identify gaps and limitations in these approaches, and trace out a path for future research that addresses the crucial issues of scale-up, and exploiting hierarchy and modularity.

## 1.1 The Statechart Model

Statecharts were introduced by David Harel in 1987 [21] as a visual formalism for complex *reactive systems* [25]: event-driven systems which continuously react to external stimuli. Examples of such systems include communication networks, operating systems, and embedded controllers for telephony, automobiles, trains and avionics systems. The primary motivation behind this model was to overcome the limitations inherent in conventional state machines in describing complex systems – proliferation of states with system size and lack of means of structuring the descriptions. The statecharts model extends state machines along three orthogonal dimensions – hierarchy, concurrency and communication – resulting in a compact visual notation that allows engineers to structure and modularise system descriptions.

1. Hierarchy is the ability to cluster states into a superstate (an OR state), or refine an abstract state into more detailed states; when a system is in a superstate, it is in *exactly one* of its sub-states. Visually the hierarchy is denoted by physical encapsulation of states (rounded rectangles).

2. Concurrency denotes orthogonal subsystems that proceed (more or less) independently and is described by an AND decomposition of states: when a system is in a composite AND state, it is in *all* of its AND components. Visually an AND state is depicted by physically splitting a state using dashed lines.

3. Communication between concurrent components is via a broadcast mechanism. Variants using directed communication along named channels is also common. Both synchronous and asynchronous styles of communication have been proposed.

All these features entail a rather complex structure on the transitions. A simple transition may have a triggering event (whose occurrence cause the transition to take place), an enabling guard condition (which must be true for the transition to be taken), and output event and actions, all of which are optional. Transitions can have OR and AND states as source and targets, can be between states at different levels of the state hierarchy and can be composed from simple transitions using fork, join and conditional connectives. In addition, OR states may have a default initial state (indicated by a small incoming arrow with a black circle at their tail) and history states to indicate that when entering an OR state, the substate entered is the most recently visited. For details the reader is referred to Harel's original paper [21] or the book by Harel and Politi [27].

**Example 1.1** *The statechart in Figure 1 is taken from [41]. The top level state `TV` is an OR state, whose substates are `WORKING` and `WAITING`. `WORKING` is an AND state whose orthogonal substates are `IMAGE` and `SOUND`, each of which is an OR state. The top level default state is `WAITING`. The transition labels `t0` through `t8` are used for ease of exposition and are not part of the statecharts syntax. The transition*

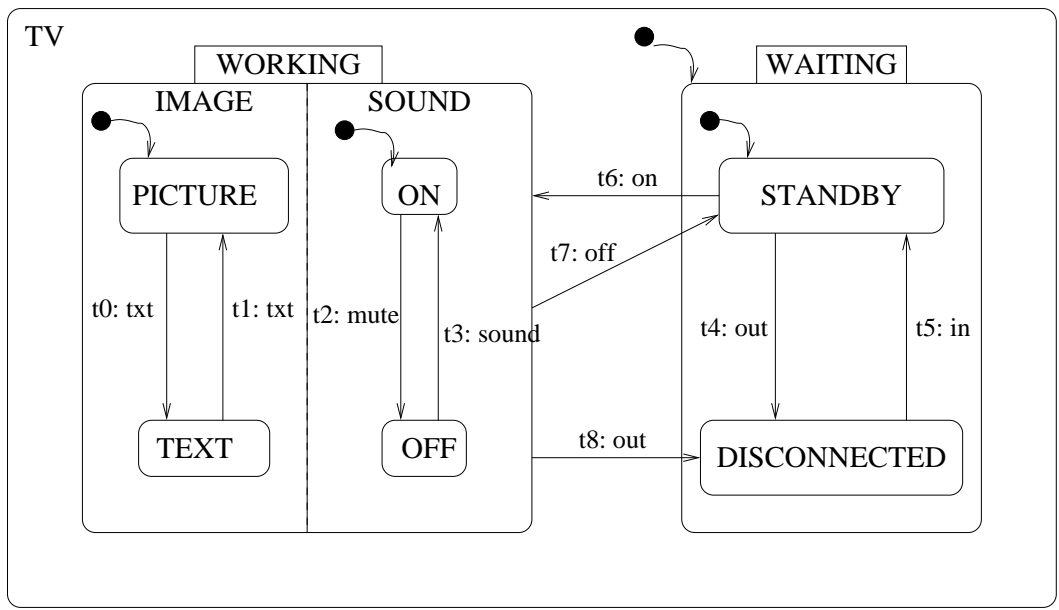

Figure 1: Example statechart

*with trigger* out *from the* AND *state* WORKING *to* DISCONNECTED *is an inter-level transition, with its source and target at different levels of the state hierarchy.*

Over the years, a number of variants of the statecharts model have been proposed – MODECHARTS [32], RSML [35], OMT [47], ROOM [48] and UML [9], to name a few. They retain the original statecharts structuring mechanism of hierarchy and group transitions, but differ significantly in semantic aspects. The most popular of these variations is the one used in the Unified Modeling Language (UML) [9], a variant of statecharts suitable for object-oriented modelling. We will describe some of these statecharts variants in subsequent sections when we discuss approaches to model checking of these models.

## 1.2 The Semantics of Statecharts

Central to the question of formal verification of statecharts is their *semantics*. Unfortunately, defining the semantics of a complex notation like statecharts is not a straightforward task. The original statecharts paper by Harel [21] only hinted informally at how a semantics could be defined. The first rigorous semantic definition was proposed in [26], followed by a series of semantic proposals (see [18, 24, 30, 31, 44] for a sampler). We briefly describe the design decisions taken by the STATEMATE semantics of statecharts, as presented in [24]. STATEMATE [23] is a CASE-tool from i-Logix for the model based development of complex reactive systems. STATEMATE uses the visual notation of statecharts for the description of control, data transformations and timing aspects of systems. The main difference between the STATEMATE semantics and the one proposed in [26, 44] is that changes that occur in a given step take effect in the next step. The semantics of a system is a set of possible *runs*, where each run is the response of the system to a sequence of external stimuli. A run can be thought of as a sequence of snapshots (status) of the system, where each status is obtained from its predecessor by executing a *step*. The move to a new status by a step is triggered by external stimuli provided by the environment at the beginning of the step, as well as changes that occurred during and since the previous step. The execution of a step itself takes zero time. The main difficulty in defining the effect of a step is in handling a set of enabled transitions possibly containing conflicting ones, and in handling inter-level transitions, i.e., those crossing the boundaries of nested states. The first step in defining an operational semantics of statecharts is to define a *configuration* as a maximal set of states that the system can be in simultaneously. From the intuition behind OR and AND states, a configuration $C$ is a nonempty upward closed (under OR hierarchy) set of states, such that for every OR state $S$ in $C$, there is exactly one substate of $S$ in $C$, and for every AND state $T$ in $C$, all its substates are in $C$. The exact details of how compound transitions, actions and conflicts are handled can be found in [24]. The salient features of the semantics are:

1. Reactions and changes that take place in a step can only be sensed in the next step.

2. Events live for the duration of one step and are lost henceforth.

3. Computations in a step are based on the status at the beginning of the step.

4. A maximal subset of nonconflicting transitions is always executed.

There are two distinct models of time in this semantics: the *synchronous* and the *asynchronous* (or *super-step*) model. In the synchronous model the system executes a single step every clock tick, reacting to all the changes that occurred in the one time unit since the previous step. In the asynchronous model, the system reacts to an external stimulus by performing a sequence of reactions (called *steps*). At each step, a maximal conflict-free set of enabled transitions is selected based on events and conditions generated in the previous step. While all events live for one step, external events are consulted only at the first step and are communicated to the environment after completion of the last step in a super-step. Super-steps are executed infinitely fast, with the clock being incremented only at super-step boundaries.

One of the most popular variants of statecharts used in software engineering practice is the state machine model in UML [9], as described in [22, 42]. This is an object-oriented variant of statecharts, which captures the dynamics of an object's internal behaviour in response to service requests from other objects and the environment. We briefly mention the differences between UML statecharts and the classical model. The mechanism for inter-object interaction includes, in addition to events, invocation of an operation on an object. When an object generates an event for another object it is queued in a system queue (in a single threaded system); when the sending object reaches a stable condition (all orthogonal components are in states and no transitions enabled due to local conditions), the system delivers the event to the appropriate object's state machine for processing. When an object calls an operation of another object, the calling object's statechart is suspended, and the callee object executes a method on its behalf. A trigger for a transition is either an event or an operation request. As in the STATEMATE semantics, reactions to events take place in steps, with events and actions generated in one step being detected in the next step, after a stable situation has been reached. Unlike the STATEMATE semantics, events are queued instead of being transmitted without delay. There is also a difference in transition priorities, with UML state machines giving priority to transitions deeper in the state hierarchy. The main difference, however, is in the *run-to-completion* (RTC) semantics of UML statecharts: all the locally enabled transitions of a statechart caused by the processing of an event will be executed before the next event will be taken for execution. For a formal treatment of the semantics refer to [17].

## 1.3 Challenges in the Verification of Statecharts Models

The graphical statecharts language offers a rich set of features for extending standard state transition diagrams with parallelism, hierarchy and broadcast or directed communication. Some of these features are: complex state hierarchies and configurations, inter-level transitions, group transitions, transition priority and simultaneous execution of maximal non-conflicting sets of transitions. These complex features interact in intricate and unexpected ways. It is a veritable challenge to provide a coherent formal semantics to these semantically rich features and faithfully implement them in a CASE tool, as witnessed by the flood of proposals for statecharts variants and their formal semantics.

The same challenges are encountered in model checking of statecharts. In the model checking framework, the system is represented as a transition system (equivalently, a Kripke structure[1] [16]). A transition system is defined as a tuple $(Q, \iota, R)$, where $Q$ is the set of *states*, usually specified by assignments of values to a set of variables $V$; $\iota$ is a set of states (expressed as predicates on $V$) defining the initial set of states; and $R$ is the transition relation, usually expressed by predicates containing unprimed and primed variables from $V$ for the pre- and post-state. Model checking of statecharts in this framework requires an interpretation of the statecharts model as a transition system. This has proved to be a significant challenge. We list below some features of statecharts that complicate this interpretation of statecharts as transition systems, making automated verification of statecharts models a difficult problem.

**State Hierarchy** The straightforward way to represent a statechart as a transition system is to flatten its hierarchy. However, this can lead to an exponential blow-up in size, particularly when there is a lot of sharing of states and transitions. While model checking of hierarchical state machines have been investigated in the literature [5, 6] using hierarchical Kripke structures, this model is not the original statecharts model: inter-level transitions, concurrent states and reading and writing of variables are not considered.

---

[1]Kripke structures have an additional component: a labelling of each state with the set of propositions that hold in that state; in a transition system, the set of propositions that hold in a state is given by the set of predicates on state variables that evaluate to true in that state.

**Non-compositionality: inter-level transitions** A syntax-directed, hierarchical semantics and translation scheme for statecharts is crucial for efficient verification of their correctness. A *compositional semantics* would interpret the meaning of a composite statechart in terms of the semantics of its constituent components, without having to consult their internal structure. Unfortunately, the statecharts model has *inter-level transitions* as a rich transfer of control mechanism across state hierarchies. The semantics of inter-level transitions violates the state encapsulation hierarchy and hence compositionality. Inter-level transitions combined with transition priorities have intricate semantics and precludes any straightforward formalisation or translation.

**Conflicting Transitions and Transition Priority** Two transitions are in conflict if there is a common state shared by their source states. If two conflicting transitions are at the same level of the state hierarchy, the result is *nondeterminism*, which is an undesirable phenomenon in the execution of a reactive system, because it implies the same sequence of inputs can lead to different output sequences in different runs. On the other hand, if the source states of one transition subsume another, i.e., if the first transition is at a higher level in the state hierarchy, then it is given *higher priority* in the STATEMATE semantics (the situation is reversed in the UML semantics). Handling nondeterminism and priority schemes are often a challenge in formal models of statecharts. Automatic detection of deadlock is also a challenge, since the enabledness of transitions depend on conditions on data values, and the problem is undecidable in general.

**Concurrency** A faithful modelling of concurrency in statecharts (either at the level of concurrent substates or multiple statecharts) requires the scheduling of concurrent execution of transitions at a fine level of granularity. Issues like multi-threaded and single-threaded execution complicate the matter. Most modelling formalisms have two principal means of handling concurrency: by synchronous execution and interleaving, with very little control in mixing the two.

**Communication** Several dimensions of communication mechanisms can be identified in statecharts variants: broadcast versus point-to-point, synchronous versus asynchronous, instantaneous versus timed. Modelling of asynchronous communication in transition systems causes subtle problems, because the levels of granularity of interleaving in statecharts may be different from the one employed by the modelling language, which in most cases is fixed. Asynchronous models also lead to a blow-up in the size of the state space, making model checking impractical for real life examples.

**History** The statecharts model has two history connectors: H (shallow history) and H$^*$ (deep history). When entering an OR state by shallow history, the substate entered is the one most recently visited; this applied to only the top level immediate substates – at deeper levels the entry is to the default states. On the other hand, on entry by deep history, the basic configuration (including all recursively contained substates) last visited relative to the OR state is entered. The modelling of deep history implies all state configurations are stored in a variable that retains its value between transitions, whereas for shallow history only the state variable corresponding to the top level substate last visited should be retained. This complicates the treatment of variables in the model, with some variables requiring special treatment.

**Models of Time** The definition of what constitutes a step is central to the semantics of statecharts. The synchronous semantics of statecharts referred to above is simpler to model using transition systems. The greediness of transition execution inherent in the asynchronous semantics makes it difficult to model. The same remark applies to the run-to-completion semantics of the UML state machine model.

## 2 Survey of Statecharts Model Checking Approaches

### 2.1 Model Checking of Statecharts by translating to SMV

SMV [39] is a model checking tool for the verification of FSMs against specification in the temporal logic CTL [16]. The input language of SMV allows the description of FSMs that range from completely synchronous to completely asynchronous. SMV uses BDDs [11] to represent state sets and transition relations and a symbolic model checking algorithm for verification. It has been applied successfully to many industrial scale hardware circuits. Its application to the analysis of software specifications has been somewhat limited.

CTL **Model Checking using** SMV  The temporal logic model checking problem is to answer the question: does the model of a (software or hardware) system, given as a finite state transition system, satisfy a given property specified in a language of temporal logic. One such logic is the branching-time temporal logic CTL [16]. Temporal logics for transition systems provide operators for describing properties of individual runs of the transition system together with mechanisms for quantifying over such runs. For example, "all execution sequences starting in state $s$ eventually lead to state $s'$." The logic CTL$^*$ defines such properties using *state formulas and* path formulas, starting from a given set of propositions (usually predicates on the state variables). A path formula is either a state formula, a Boolean combination of path formulas, X $P$ ($P$ holds in the next-state) or $P$ U $Q$ ($P$ holds until $Q$ holds) where $P$ and $Q$ are path formulas. Derived path formulas such as F $P \equiv$ TRUE U $P$ (eventually $P$) and G $P \equiv \neg F(\neg P)$ (always $P$) are frequently used. A state formula is either a proposition, any Boolean combination of state formulas, or A $P$ ($P$ holds along all paths), where $P$ is a path formula. The derived state formula E $P \equiv \neg$A $(\neg P)$ (there exists a path along which $P$ holds) is also used frequently. CTL formulas are state formulas where every path modality (X,U,F or G) is immediately preceded by a path quantifier (A or E). Examples of CTL formulas are:

AG($\neg cs_1 \vee \neg cs_2$): In all reachable states, processes $p_1$ and $p_2$ are not both in their
critical section (where $cs_i$ indicates process $i$ is in its critical section).

AG AF *stable*: The system is stable infinitely often.

AG($sent \rightarrow$ AF $received$): Every message sent is received.

AG EF *reset*: It is possible to *reset* the system in any reachable state.

A transition system $T = (S, \iota, R)$ satisfies a formula $\phi$ if $\phi$ holds in all initial states in $\phi$. A model checker tries to verify that $T$ satisfies $\phi$ by an exhaustive but efficient state space search. If the formula does not hold, the model checker returns a counterexample trace leading to a state where the formula is violated. A model checker based on *explicit state enumeration* verifies the truth of a CTL formula by traversing the state diagram in a depth-first or breadth-first manner. The time complexity is linear in the state space and the length of the formula [15]. However, the state space grows exponentially with the number of system variables and components, leading to the *state explosion problem*. IN the late eighties an important advance in verification technology took place with the invention of *symbolic model checking* [39], the transition relation is not explicitly constructed but represented implicitly by a Boolean function. Sets of states are also represented by Boolean functions. Boolean functions are succinctly represented using BDDs [11], which are efficient ways of representing Boolean functions using sharing of subexpressions and elimination of redundant variables in binary decision trees. Instead of visiting individual states as in explicit state space search, symbolic model checking relies on visiting a *sets* of states at a time, using efficient BDD representations of both states and transition relations.

SMV [39] is a symbolic model checker for properties expressed in CTL. An SMV model of a system consists of a finite-state system description and a list of CTL formulas. The state space is defined by state variable declarations, for example:

```
VAR
state0: {noncritical, trying, critical};
state1: {noncritical, trying, critical};
turn: boolean;
```

The transition relation and the initial states are specified by a collection of simultaneous assignments to state variables:

```
ASSIGN

init(state0) := noncritical;

next(state0) :=
case
   (state0 = noncritical) : {trying,noncritical};
   (state0 = trying) & (state1 = noncritical): critical;
   (state0 = trying) & (state1 = trying) & (turn = turn0):  critical;
   (state0 = critical) : {critical,noncritical};
```

```
      1: state0;
esac;

init(turn) := 0;
next(turn) :=
case
    turn = turn0 & state0 = critical: !turn;
    1: turn;
esac;
```

For any variable v, `init(v)` refers to the initial value of v and `next(v)` refers to the value of v in the next state. Next state values of variables are often defined as parts of a `case` expression, as in the above example. In SMV, 1 and 0 represent TRUE and FALSE respectively. The logical operators AND, OR and NOT are represented by `&`, `|` and `!`, respectively. It is important to note that the assignments to the variables `state0` and `turn` are concurrent: all state variables in an SMV module are assigned new values simultaneously at the beginning of a new transition. An alternative way to specify the transition relation between states is to use propositional formulas relating the old and new states:

```
 INIT
    output = 0
 TRANS
    next(output) = !input | next(output) = output
```

SMV also allows the use of macro definitions to abbreviate expressions:

```
DEFINE
  carry_out := value & carry_in;
```

These definitions do not add variables to the state space but are convenient shortcuts that are expanded when generating the state machine.

SMV allows the definition of reusable modules and expressions:

```
MODULE counter_cell(carry_in)
 VAR
    value : boolean;
 ASSIGN
    init(value) := 0;
    next(value) := value + carry_in mod 2;
 DEFINE
  carry_out := value & carry_in;

MODULE main
 VAR
    bit0 : counter_cell(1);
    bit1 : counter_cell(bit0.carry_out);
```

```
   bit2 : counter_cell(bit1.carry_out);
 SPEC
   AG AF bit2.carry_out
```

The examples above describe synchronous systems, where the assignment statements are executed simultaneously. SMV can also model asynchronous systems by defining a collection of parallel processes, whose actions are interleaved:

```
MODULE inverter(input)
 VAR
   output : boolean;
 ASSIGN
   init(output) := 0;
   next(output) := !input;
 FAIRNESS
   running

MODULE main
 VAR
   gate1 : process inverter(gate3.output);
   gate2 : process inverter(gate1.output);
   gate3 : process inverter(gate2.output);

SPEC
  (AG AF gate1.out) & (AG AF !gate1.out)
```

The example above describes a ring of three asynchronous inverting gates. Asynchronous processes are declared by the keyword `process` before instantiating the module. At a given instant only one process is chosen nondeterministically for execution, and all its assignments executed in parallel. The `FAIRNESS running` constraint forces every instance of inverter to execute infinitely often.

### 2.1.1 RSML Model Checking [Chan et al 1998]

One of the first applications of SMV to the analysis of statecharts models was in [12], where the authors use RSML [35], a variation of Harel statecharts, as the input model. RSML borrows the notions of superstates, AND decomposition and broadcast communication from statecharts. It adds features like interface description and directed communication between state machines, in addition to some syntactic variations for specifying guards and conditions. The RSML models were translated to SMV and the authors claim that they were able to control the size of the BDDs representing the specification to analyse a number of robustness and safety-critical properties. The salient features of their translation scheme are:

1. SMV variables are introduced for state hierarchy, inputs and events, as follows:

   - Events translate to boolean variables.

- Input variable translate to variables with an enumerated or subrange type.
- Each OR-state translates to an enumerated variable which ranges over the substates of the OR-state. The values of these variables completely determine the current configuration of the state machine.

2. A defined boolean symbol indicates the condition under which the machine is in a particular state.

3. A defined boolean symbol for each transition expresses when the transition is enabled – when the machine is in the source state, the trigger event occurs and the guarding condition is true.

4. The state change is specified using the `next` state update of SMV inside an `ASSIGN` clause. An inner `case` statement ranging over all the transitions specifies the next state and another `next` statement specifies the event generated.

5. States and events are initialised using an `init` clause.

6. Inputs to the machine are modelled non-deterministically to allow arbitrary environmental behaviour while satisfying the synchrony hypothesis: inputs do not change when the machine is not stable.

**Example 2.1** *The SMV program for the statechart in Figure 1 is shown below:*

```
MODULE main
VAR
-- events
  txt: boolean;
  mute: boolean;
  sound: boolean;
  out: boolean;
  in: boolean;
  on: boolean;
  off: boolean;
-- OR states
  TV: {WORKING, WAITING};
  IMAGE: {PICTURE, TEXT};
  SOUND: {ON, OFF};
  WAITING: {STANDBY, DISCONNECTED};
DEFINE
-- state machine is in stable state
  stable := !(txt | mute | sound | out | in | on | off)
-- condition under which machine is in a particular state
  in-TV := 1;
  in-WORKING := in-TV & TV = WORKING;
  in-WAITING := in-TV & TV = WAITING;
  in-IMAGE := in-WORKING;
  in-SOUND := in-WORKING;
```

```
    in-PICTURE := in-IMAGE & IMAGE = PICTURE;
    in-TEXT := in-IMAGE & IMAGE = TEXT;
    in-ON := in-SOUND & SOUND = ON;
    in-OFF := in-SOUND & SOUND = OFF;
    in-STANDBY := in-WAITING & WAITING = STANDBY;
    in-DISCONNECTED := in-WAITING & WAITING = DISCONNECTED;
-- enabledness of transitions
    t0 := in-PICTURE & txt;
    t1 := in-TEXT & txt;
    t2 := in-ON & mute;
    t3 := in-OFF & sound;
    t4 := in-STANDBY & out;
    t5 := in-DISCONNECTED & in;
    t6 := in-STANDBY & on;
    t7 := in-WORKING & off;
    t8 := in-WORKING & out;

ASSIGN
    init(TV) := WAITING;
    next(TV) :=
      case
        t6: WORKING;
        t7: WAITING;
        t8: WAITING;
         1: TV;
      esac;

    init(IMAGE) := PICTURE; -- default state
    next(IMAGE) :=
      case
        t0: TEXT;
        t1: PICTURE;
        t6: PICTURE;   -- entry to default state
         1: IMAGE;
      esac;

    init(SOUND) := ON;
    next(SOUND) :=
      case
        t2: OFF;
        t3: ON;
        t6: ON;  -- entry to default state
         1: SOUND;
      esac;

    init(WAITING) := STANDBY;
```

```
  next(WAITING) :=
    case
      t4: DISCONNECTED;
      t5: STANDBY;
      t7: STANDBY;
      t8: DISCONNECTED;
       1: WAITING;
    esac;

-- events
  next(txt) :=
    case
      stable: {0,1}
           1: 0;
    esac;
 .
 .
 .
```

*Note that all the events in this example are generated by the environment, and hence they are nondeterministic, except they are not allowed to change when the machine is not stable. If output events are generated by the machine, their handling similar to the handling of state changes by the* next *statement.*

**Discussion**

- **Nondeterminism** The translation scheme presented above works only for deterministic machines. Since SMV has a concurrent evaluation semantics in which all transitions enabled by a given input event fire simultaneously, the translation does not preserve the semantics of nondeterministic statecharts. Nondeterminism can be handled in the following way. Instead of representing transitions as defined symbols, they should be declared as ordinary Boolean variables. Then the enabledness of transitions can be represented as a first-order logic formula over the finitely many auxiliary variables and the global state variables. This method is inefficient because the number of transitions is usually large.

- **Synchrony Hypothesis** The RSML semantics of statecharts is similar to the STATE-MATE semantics discussed above, except for the synchrony hypothesis in the former. The synchrony hypothesis can be dropped by defining the symbol stable to 1.

- **Transition Priority** RSML does not use the priority scheme for transitions used by STATEMATE to resolve certain conflicting transitions. To interpret priority, certain modifications in the transition rules are necessary.

- **Other features** RSML does not have history connectors, synchronisation through activities and optional trigger events. In principle, these features can be handled by new translation rules.

### 2.1.2 Translation of STATEMATE statecharts to SMV [Clarke and Heinle, 2000]

The STP-approach to translating STATEMATE statecharts to SMV, reported in [14], attempts to reflect the hierarchical structuring of statecharts as closely as possible in SMV to achieve a modular translation. This is in contrast to the translation by Chan et al [12] discussed in the previous section. The notion of hierarchy in this context entails that any state may contain entire statecharts (called *subcharts*), and the translation of a state depends only on the subchart it denotes. One implication of this is that inter-level transitions are not handled by the translation. The translation to SMV is via a temporal language ETL which formalises the input language of SMV. The salient features of the translation are listed below.

1. Statecharts are translated as individual SMV modules. The top level statechart is represented directly in the `MAIN` module with declarations for all global variables for events and conditions.

2. The module representing a direct subchart of a statechart is instantiated in the module corresponding to the parent statechart. The translation thus follows a hierarchical structure of statecharts.

3. Three types of SMV modules arise in the translation: the *main*-module, *chart*-modules containing translations of subcharts, and *monitor*-modules which handle global event and condition variables.

4. The following ETL predicates are used in the formalisation of statechart operations. The predicate $chart(s) = sc$ is true if the state $s$ is contained at the top level of statechart $sc$. If $s'$ is a direct subchart contained in the state $s$, this is denoted by $subchart(s', s)$. The predicate $state(sc) = s$ denotes $chart(s) = sc$ and control of the statechart is in $s$. The predicates $active(sc)$ and $active(s)$ are used to represent when control is in the statechart $sc$ and is in the state $s$ respectively.

5. The state of each statechart is modelled with a local variable `state` which ranges over values in the respective state set.

6. Initialisation and activity are communicated to the module corresponding to a subchart via two parameters `default` and `active`.

7. OR states have only one subchart whereas AND states have at least two subcharts.

8. State transitions are modelled by a `TRANS` predicate relating the current and the next state in SMV. This has the advantage over using an `ASSIGN` as in the translation of Chan et al in that it represents a nondeterministic choice between conflicting transitions in a statechart more faithfully.

9. Events are represented by global boolean variables. An event variable is set the moment the event happens and reset automatically in the next step, unless it is generated again. On the other hand, condition variables persist until they are explicitly modified. Both event and condition variables are represented by monitor modules to manipulate the corresponding global variable.

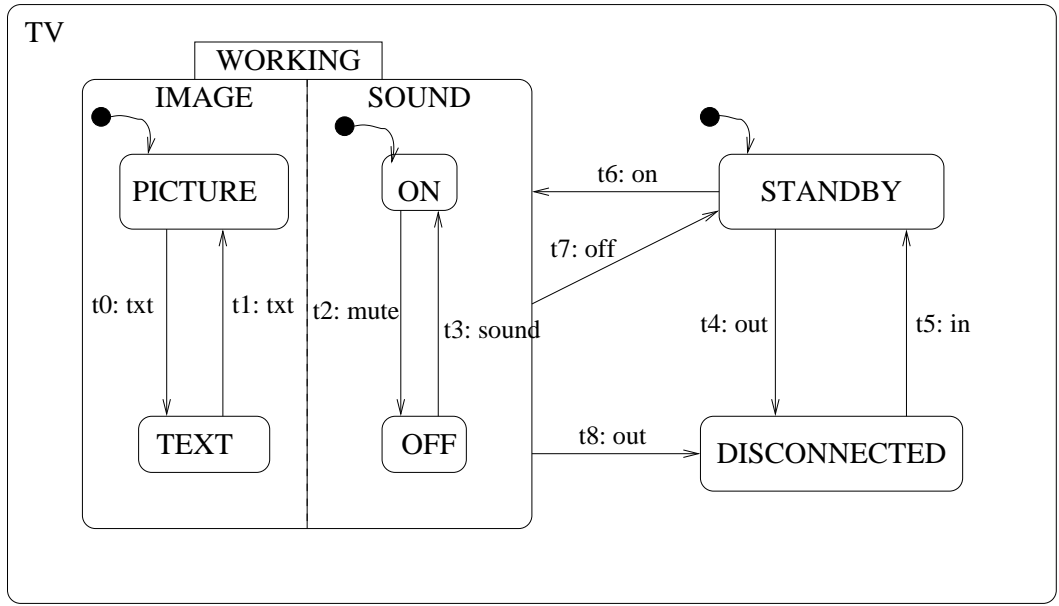

Figure 2: Statechart of Figure 1 without inter-level transitions

10. Actions and event generation are handled by a `next` within an `ASSIGN` statement.

**Example 2.2** *The statechart in Figure 2 is a modified version of the one in Figure 1, without the top-level state* `WAITING`. *Note that this removes all inter-level transitions, which the current approach does not handle. Part of the translated* SMV *code for the statechart* TV *is shown below, assuming that* TV *is not the top level state, i.e.,* TV *is contained in a bigger statechart.*

```
MODULE TV(on,off,out,in,active,default)
  VAR
    state: {WORKING, STANDBY, DISCONNECTED};
  INIT
    active & default -> state = STANDBY;
  TRANS
    next(active) & next(default) -> next(state = STANDBY);
  TRANS
    ( (active & state = WORKING & (  (out & next(state) = DISCONNECTED)
                                   | (off &next(state) = STANDBY)
                                   | (!(out|off) & next(state) = state)))
     | (active & state = STANDBY & (  (on & next(state) = WORKING)
                                   | (out &next(state) = DISCONNECTED)
                                   | (!(on|out) & next(state) = state)))
     | (active & state = DISCONNECTED & (  (in & next(state) = STANDBY)
```

```
                                                        | (!in & next(state) = state)))
          | !active)
.
.
.
```

**Discussion**

- **Events and Condition Variables** The translation is actually more involved than the above example suggests. An events is represented as a global Boolean variable, which is set at the instant the event happens and automatically reset one time step later. The problem with this scheme is that an event can be generated by any module, which may involve multiple assignments to the same variables. This is not allowed by SMV. To overcome this hurdle, events and condition variables are represented by using a *monitor* module, a concept borrowed from concurrent programming.

- **Modularity** Modules are only syntactically convenient mechanisms that make the SMV code more readable. The internal representation in SMV is a flattened out transition system.

- OR **hierarchy** There is no subroutine style hierarchical composition of modules in SMV – when a module is active, all its submodules are. This makes the SMV module construct suitable for Modeling the statecharts AND-hierarchy but not the OR-hierarchy.

- **Inter-level transitions and Priority** The modular translation of statecharts comes at a price. AS pointed out above, inter-level transitions cannot be handled, nor can the STATEMATE priority scheme for conflicting transitions.

- **State Explosion** It is not clear to what extent the state explosion problem can be avoided through the BDD based symbolic model checking procedure that SMV uses. This remark also applies to the translation of Chan et al, and indeed, any translation scheme that follows the SMV route.

## 2.2 Model Checking of Statecharts by translating to SPIN

In contrast to SMV, SPIN [29] is a model checker for the verification of asynchronous processes. SPIN uses Linear Temporal Logic (LTL) rather than CTL, for specifying the correctness properties. Further, SPIN uses an on-the-fly explicit state model checking rather than the symbolic method employed by SMV. SPIN uses a number of optimisation techniques to reduce the size of the state space, including partial order reduction. PROMELA, the input language of SPIN, is a simple program like notation for specifying process interactions using Dijkstra's guarded commands and Hoare's CSP. Process interactions can be specified with rendezvous primitives, asynchronous message passing through buffered channels, shared variables or combinations of these. SPIN has been used extensively for verifying communication protocols and distributed systems.

**LTL model checking using** SPIN   Linear-time temporal logic, or LTL [37], is a fragment of CTL$^*$ in which path formulas contain no occurrences of the path quantifiers A and E. A system satisfies the LTL formula $P$ if it satisfies the the CTL$^*$ formula A $P$. In LTL, the modalities F and G are usually written as $\Diamond$ and $\Box$, e.g., $\Diamond\Box P$ for FG $P$, "eventually always $P$".

Model checking of LTL specifications in SPIN is based on the automata theoretic approach of Vardi and Wolper [52]. The inputs to the SPIN model checker are a description of a concurrent system in PROMELA and its correctness properties expressed in LTL. The PROMELA description consists of user-defined process templates (using `proctype` definitions) and at least one process instantiation (using the `run` command). SPIN translates each process template into a finite automaton. The global behaviour of the concurrent system is obtained by computing the asynchronous interleaving product of the automata corresponding to each process. To perform verification, SPIN also converts the correctness claim in LTL to a Büchi automaton [51], and computes the *synchronous* product of the claim and the automaton for the global behaviour. If the language accepted by the resulting Büchi automaton is empty the original claim does not hold on the original system. SPIN actually uses the *negation* of the correctness claim as the input, so a non-empty intersection gives counter-examples to the correctness claim. A Büchi automaton accepts a system run iff it forces the automaton to pass through one or more of its accepting states infinitely often. Such accepting behaviours of a Büchi automaton are called its *acceptance* cycles. To prove that no execution of the system satisfies the negated correctness claim, it suffices to prove that the synchronous product of the system and the (Büchi automaton representing the) negated claim has no acceptance cycles. SPIN does the computation of automata for concurrent components, their asynchronous product representing the global system, the Büchi automaton for the correctness claim, in an on-the-fly way using a nested depth-first search algorithm (see [29] for details.).

The language PROMELA (Process or Protocol Meta Language) allows for the dynamic creation of concurrent processes. Communication via message channels can be defined to be synchronous (i.e., rendezvous), or asynchronous (i.e., buffered). We illustrate the use of PROMELA through a version of the alternating bit protocol from [28].

```
mtype = { msg, ack };

chan to_sndr = [2] of { mtype, bit };
chan to_rcvr = [2] of { mtype, bit };

active proctype Sender()
{ bool seq_out, seq_in;

/* obtain first message */
do
:: to_rcvr!msg(seq_out) ->
to_sndr?ack(seq_in);
if
:: seq_in == seq_out ->
```

```
/* obtain new message */
seq_out = 1 - seq_out;
:: else
fi
od
}

active proctype Receiver()
{ bool seq_in;

do
:: to_rcvr?msg(seq_in) ->
to_sndr!ack(seq_in)
:: timeout ->/* recover from msg loss */
to_sndr!ack(seq_in)
od
}
```

PROMELA allows message type definitions using `mtype` statement to declare symbolic values tp abstract from the specific values to be used in message passing. Message channels are used to model the transfer of data from one process to another. They are declared either locally or globally using the `chan` statement with the size of the channel in square brackets and a list of message types in braces. The `proctype` statement declares a process with parameters, but it does not run them. Such a process is instantiated by a `run` operation, which can also specify actual parameters. Alternatively, the `active` modifier can be used to make an instance of the proctype to be active in the initial system state. For message passing syntax, PROMELA uses `ch!expr` to send the value of expression `expr` to the channel `ch`, and `ch?msg` to receive the message. The message is retrieved from the head of the channel, and stored in the variable `msg`. The channels pass messages in first-in-first-out order.

The basic control flow constructs in PROMELA are *case section* using `if...fi`, and *repetition* using `do ...od` constructs, which use the syntax of guarded commands. However, the semantics of the selection and repetition statements in PROMELA are different from other guarded command languages. First, the communication can be either CSP style rendezvous or asynchronous. Moreover, the statements are not aborted when all guards are false but they block, providing the required synchronisation. In PROMELA there is no difference between conditions and statements – the execution of every statement is conditional on its executability . Statements are either executable or blocked (FALSE). The executability is the basic means of synchronisation. A process can wait for an event to happen by waiting for a statement to become executable (TRUE). PROMELA accepts two different statement separators: an arrow `->` and the semicolon `;`. The two statement separators are equivalent. The arrow is sometimes used as an informal way to indicate a causal relation between two statements.

The `timeout` statement models a special condition that allows a process to abort waiting for a condition that may never become true. It provides an escape from a deadlocked or hang state. The timeout condition becomes true only when no other statements within the distributed system is executable.

The following example illustrates the implementation of a Dijkstra semaphore, using binary rendezvous communication. This is achieved by declaring sema to be a channel of size 0.

```
#define p 0
#define v 1

chan sema = [0] of { bit };

proctype dijkstra()
{ byte count = 1;

do
:: (count == 1) ->
sema!p; count = 0
:: (count == 0) ->
sema?v; count = 1
od
}

proctype user()
{ do
:: sema?p;
   /*      critical section */
   sema!v;
   /* non-critical section */
od
}

init
{ run dijkstra();
run user();
run user();
run user()
}
```

A system described in PROMELA can be automatically analysed for correctness violations. The following types of violations are typical.

**Assertions** The statement assert(exp) statement has no effect if the boolean condition exp holds. If the condition does not necessarily hold, i.e., there is an execution sequence in which the condition is violated, the statement will produce an error report during verifications with Spin.

**End-states** Valid end-states are those system states in which every process instance and the init process has either reached the end of its defining program body or is blocked at a statement that is labelled with a label that starts with the prefix `end`. All other states are invalid end-states, signifying deadlocks. During verification an error is reported if there is an execution that terminates in an invalid end-state.

**Progress states** A progress state is any system state in which some process instance is at a statement with a label that starts with the prefix `progress`. A *non-progress cycle* is detected by the verifier if there is an execution that does not visit a progress state infinitely often. Non-progress cycles indicate the possibility of starvation or lock-out.

**Acceptance states** An acceptance state is any system state in which some process instance is at a statement with a label that starts with the prefix `accept`. An error is reported by the verifier if there is an execution that visits an acceptance state infinitely often.

**Temporal claims** LTL formulae can be used to express general safety and liveness properties. SPIN compiles an LTL formulae into a *never claim*, the negation of the correctness property. A never claim `never {statements}` is a special type of process that, if present, is instantiated once. It is equivalent to a Büchi automaton representing the negated property, and is used to detect behaviours that are considered undesirable or illegal. When checking for state properties, such as assertions, the verifier reports an error if there is an execution that ends in a state in which the never claim has terminated. i.e., has reached the end of its body. When checking for acceptance cycles, the verifier reports an error if there is an execution that visits infinitely often an acceptance state. Thus, a temporal claim can detect illegal infinite (hence cyclic) behaviour by labelling some statements in the never claim with an acceptance label. In such situations the never claim is said to be matched. In the absence of acceptance labels, no cyclic behaviour can be matched by a temporal claim. Also, to check a cyclic temporal claim, acceptance labels should only occur within the claim and nowhere else in the PROMELA system. A never claim is intended to monitor every execution step in the rest of the system for illegal behaviour and for this reason it executes in lock-step with the other processes (synchronous product). Such illegal behaviour is detected if the never claim matches along a computation. If a claim blocks (because no statement in its body is enabled) but it is not at the end of its body then there is no need to explore this computation any further.

### 2.2.1 Implementing STATEMATE statecharts in SPIN [Mikk et al, 1998]

The work reported in [41] uses the extended hierarchical automata (EHA) model of [40] as an intermediate format for translating STATEMATE statecharts to SPIN. The choice of EHA was motivated by the need for a structural operational semantics definition for statecharts which is difficult in the presence of interlevel transitions. The EHA model uses transitions between states at the same level by lifting interlevel transitions to the uppermost states that are exited and entered, with annotations on the transitions to describe the actual source and target.

**Definition 2.1** *An EHA consists of a set of sequential automata. A* sequential automaton *A is a 4-tuple* $(\Sigma, s_0, L, \delta)$ *where $\Sigma$ is the set of states, $s_0 \in \Sigma$ is the initial state of A, L is the set of transition labels and $\delta \subseteq \Sigma \times L \times \Sigma$ is the transition relation. For a set of sequential automata $F = \{A_1, \ldots, A_n\}$ with mutually disjoint state spaces, the* composition function $\gamma : \bigcup_{A \in F} \Sigma_A \to \wp(F)$ *maps a state s of a sequential automaton to a set of automata $G \subseteq F$; s is said to be* refined *by G in such a case. A composition function is required to be tree-like with designated root automaton $\gamma_{root}$. If $|\gamma(s)| = 1$ then s is refined by a single automaton; if $|\gamma(s)| > 1$ then s is refined into a parallel composition of automata. If $\gamma(s) = 0$, s is a* basic state.

As in usual in defining the semantics of statecharts, system states of an EHA $H$ are modelled by configurations. A *configuration* is a set of states of the component sequential automata of $H$, with every sequential automaton contributing at most one state to a configuration. The root automaton $\gamma_{root}$ is part of every configuration. Whenever a non-basic state is in a configuration, each of its direct sub-automata must contribute to the configuration, and vice versa. The set of all configurations is denoted $Conf(\gamma)$. The initial configuration is derived from he initial states of the set of sequential automata in a top-down manner starting from the root automaton.

The label $L$ of a transition from source state $s$ to target state $s'$ is a 4-tuple $(sr, ex, ac, td)$:

- The *source restriction* $sr \in Conf(\gamma \restriction s)$ of a label is used to restrict the enabledness of the transition to a set of sub-configurations below $s$.

- The *transition guard* is a proposition $ex$ over events and state names. The set of events of an EHA is denoted $E$. Models of $ex$ are statuses: pairs (C,E) where C is a configuration and E is a set of events.

- The set of generated events is $ac$.

- The *target determinator* $td \in Conf(\gamma \restriction s')$ is used to determine the states to be entered in the sub-automata of the transition target.

A transition $t = (s, (sr, ex, ac, td), s')$ of an automaton $A \in F$ is *enabled* in a status (C,E) where C$\in Conf(\gamma)$ and E $\subseteq E$, iff the source state is active i.e., $s \in$ C, the source restriction is an active sub-configuration $sr \subseteq$ C, and transition guard $ex$ is enabled, (C,E) $\models ex$.

**Example 2.3** *The EHA corresponding to the statechart in Figure 1 is shown in Figure 3. The original statechart is transformed to sequential automata* TV, IMAGE, SOUND *and* POWER, *depicted by dashed rounded boxes. The state* WORKING *of sequential automaton* TV *is refined into the set of sequential automata* {IMAGE,SOUND}, *denoting their parallel composition. The state* WAITING *is refined into the singleton set* {POWER}. *The interlevel transitions labelled* t6, t7 *and* t8 *in Figrefexample-1 are replaced by transitions labelled* l3, l1 *and* l2, *respectively. Note that in contrast to statecharts, a transition in an EHA always resides within one sequential automaton. The transition labelled* l3 *is enabled if* WAITING *and* STANDBY *are active and the event* on *is present. The effect of taking the transition is that the states* WORKING, PICTURE *and* ON *are entered.*

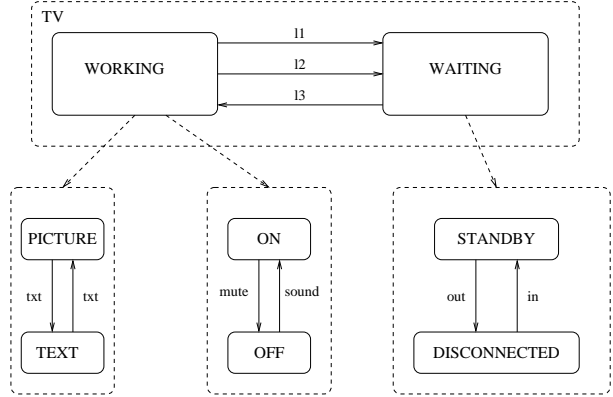

| Label | Source restriction | Guard | Target determinator |
|-------|--------------------|-------|---------------------|
| l1    | ∅                  | off   | {STANDBY}           |
| l2    | ∅                  | out   | {DISCONNECTED}      |
| l2    | {STANDBY}          | on    | {DISCONNECTED}      |

Figure 3: Extended Hierarchical Automaton for Figure 1

The translation from statecharts to SPIN is based on the formal operational semantics of STATEMATE as presented in [40], based on the informal presentation in [24]. The semantics of an EHA $H = (F, E, \gamma)$ is given in terms of a Kripke structure $\mathbf{K} = (\mathbf{S}, \mathbf{s_0}, \stackrel{STEP}{\rightarrow})$, where $\mathbf{S} = Conf(\gamma) \times \wp(E)$ is the set of states (or statuses) of $\mathbf{K}$, $\mathbf{s_0} = (C_0, \emptyset)$ is the initial state of $\mathbf{K}$ and $\stackrel{STEP}{\rightarrow} \subseteq \mathbf{S} \times \mathbf{S}$ is the transition relation of $\mathbf{K}$ defining the operational semantics of EHA. The operational semantics, presented in SOS style, is in terms of three principal rules:

**Progress Rule**: This rule applied to a sequential automaton $A$ if one of its states $s$ is in configuration C and if one of the outgoing transitions is enabled; one of them is taken non-deterministically. The transition label determines the effect of the transition: the target state $s'$ and the target determinator states are entered, and events of $ac$ are generated.

**Composition Rule**: This rule applies to an automaton $A$ that has one of its states $s$ in the configuration C but all outgoing transitions are disabled. If the state $s$ is refined to a set of automata $\{A_1, \ldots, A_n\}$, the rule delegates the step to the sub-automata by collecting the results of the steps performed by the $A_i$.

**Stuttering Rule**: This rule applies to a basic state $s$ in the configuration C with none of its outgoing transitions enabled. The effect is to remain in state $s$ without generating any events.

The details of the operational semantics can be found in [40, 41].

The paper presents two translation frameworks from EHA to SPIN, leading to sequential or parallel code, of which the former can be verified more efficiently. The sub-language of statecharts considered in the paper restricts the set of actions to only generated events and does not consider data transformations, history and timing issues.

Given an EHA and its operational semantics in terms of a Kripke structure $\mathbf{K}$, the translation maps the EHA to a PROMELA model $P$. The salient features of the translation are listed below.

1. The PROMELA model $P$ contains variables necessary to encode states of $\mathbf{K}$ and a PROMELA process or processes encoding the transition relation of $\mathbf{K}$.

2. The operational semantics of STATEMATE statecharts in terms of EHA is used to achieve a structural translation – rules of the semantics are used as generic pattern while generating code for EHA.

3. Configurations of an EHA are implemented by defining for every sequential automata with $n$ states a variable that distinguishes at least $n + 1$ different values – $n$ values represent the states of the automaton and the extra value to model that the automaton does not have an active state in the current configuration.

4. Events are represented by boolean variables.

5. Because parallel composition in PROMELA has an interleaving rather than synchronous semantics, to implement concurrent access to the variables encoding states and events, two copies of each variable are needed to encode the pre- and post-state.

6. The translation supports closed systems only, i.e., the specification should contain an abstract model of the environment.

7. To implement the behaviour of AND states, a parallel composition of EHA must be implemented. The challenge is to implement it in PROMELA, which uses an interleaving model of parallelism. There are two choices for this implementation:

    - Parallel solution. Transitions in parallel automata are executed in an interleaved fashion, i.e., parallel automata are mapped to PROMELA processes. This requires a scheduler to implement the transitions in a "lock-step" manner, as dictated by the semantics.

    - Sequential solution. An arbitrary interleaving of parallel transitions is determined at compile-time and translated to a fully sequential implementation.

8. The operational semantics of STATEMATE statecharts in terms of EHA induces a depth-first search traversal strategy of the hierarchical structure of EHA. First the root automaton is considered for enabled transitions. If it does not have an enabled transition in the active state then the sub-automata of the active state are scheduled, control being passed to each of them in arbitrary order. This translation schema has two advantages:

- Hierarchical structure is mapped to if-clauses of PROMELA, faithfully implementing the hierarchy of STATEMATE statecharts.

- Redundant checks are avoided by checking higher-level transitions for enabledness first.

9. The implementation of a sequential automaton either executes one of its transitions, depends on the execution of its subcomponents or remains in its current state. All these cases are implemented as conditions over the pre-state variables and assignments to post-state variables.

10. In order to generate infinite runs of the PROMELA model the code for the transition relation is wrapped into an infinite loop.

**Discussion**

**Statecharts subset** The translation handles a limited subset of STATEMATE statecharts - no data transformations (such as actions with assignments to variables), timing primitives or history are supported. In principle, some of these limitations can be handled by adding more features to the EHA semantic model and corresponding translation rules.

**Model extractor and SPIN back end for error reporting** The translator automatically translates graphical STATEMATE models into PROMELA with a SPIN back-end that translates counterexamples back into terms of the original specification. Thus the translation to PROMELA is hidden from the user.

### 2.2.2 Translating UML Statecharts to PROMELA/SPIN [Latella et al, 1999]

The paper by Latella et al [33] proposes a translation from a subset of UML statecharts into PROMELA. The work is based on an operational semantics of UML statecharts presented in [34] which is similar to the approach in [40], the basis for the work by Mikk et al described in the previous section. Like the previous work, the subset of statecharts considered does not include history, action and activity states. Further, events are restricted to signal and call events, without parameters; time and change events, object creation and destruction events and deferred events and branch transitions are not not considered. Variables and data are not considered so that actions can only generate events. Entry and exit actions of states are also abstracted away. The translation to PROMELA follows the basic approach of Mikk et al, with appropriate modifications to account for UML statecharts semantics. Some of these considerations are the complications induced by the UML transition priorities and their reverse relation with the hierarchical structure of states. The translation is parametric with respect to a notion of priority schema, which can be instantiated with the UML specific case. It is also parametric with respect to the environment, unlike the previous work, where the environment is represented as a set of events at each step. The authors claim that the code generated is also much simpler and does not need to use pre- and post-variables.

Here are the important features of the translation from EHA to SPIN:

1. Events are treated as uninterpreted symbols and represented as integer values.

2. Since the UML semantics of statecharts do not specify the semantics of queues for storing events directed to an object, the specifier is free to choose among a set, a multi-set and a FIFO queue, as the representation. The choice is an input parameter for the translator.

3. Sets and multisets are represented by their characteristic function ($n$ one-bit variables) and multiplicity functions ($n$ integer variables), respectively. A FIFO queue is directly mapped to a PROMELA channel whose length LT is specified by the designer.

4. An individual state is modelled by a single bit variable. A configuration corresponds to those states whose bits are set to 1.

5. The steps of the Kripke structure corresponding to the EHA are generated by the PROMELA process called STEP, which has the following four phases:

    (a) selection of an event from the environment;

    (b) identification of all the candidate transitions for firing; this includes identification of enabled transitions and resolution of conflicts based on transition priority;

    (c) selection of those transitions among the candidate ones that will be fired; this includes selection among concurrent (orthogonal substates) and choice among nondeterministic alternatives;

    (d) actual firing of the selected transitions, including identification of the resulting configuration and generation of new events.

6. The STEP process includes a loop to generate successive steps of the EHA. The atomicity of each step is ensured by using the atomic directive in PROMELA. This implies that the only values available for verification are the ones obtained at the end of each cycle.

7. The PROMELA code generated for selecting an event from the input queue, in case the queue is represented by a set, uses the selection command:

   ```
   if
   :: Qe_1 -> Ev = e_1; Qe_1 = 0
   .
   .
   .
   :: Qe_n -> Ev = e_n; Qe_n = 0
   fi
   ```

   Here Qe_i is the bit representing the presence of event e_i. In case a multi-set representation is used, the guard is Qe_i >= 1 and the action is Qe_i--. If a channel is used, the input command Q?Ev is used.

8. In order to identify the candidate transitions for firing, a boolean variable Cand_i corresponding to the transition t_i in the EHA is used. An assignment to Cand_i corresponds to implementing the progress rule in the last section.

9. The actual transition selection phase involves resolution of conflicts and selection of one transition among the candidates. This is done by nondeterministically assigning 1 to the bit variable Sel_i if Cand_i holds. The code for such an assignment is generated recursively following the tree structure of the EHA.

10. The actual firing of the selected transition t_i involves setting the bit variables for the states that are entered and resetting the variables for states that are exited. In addition, all generated events have to stored in the input queue.

**Example 2.4** *An optimised version of the* PROMELA *code generated from the statechart in Figure 1 by this approach is shown below. Note that the translation reflects the UML semantics of transition priority and concurrent execution, which is different from the intended semantics in the original statechart.*

```
/* events */
#define txt   1
#define mute  2
#define sound 3
#define on    4
#define off   5
#define out   6
#define in    7

/*states */
bit WORKING, PICTURE, TEXT, ON, OFF, WAITING, STANDBY, DISCONNECTED
/* set of events in the current environment */
bit Q_txt, Q_mute, Q_sound, Q_on, Q_off, Q_out, Q_in;
/* selected event */
int Ev;

/* whether transition t_i is a candidate for firing */
#define Cand_0 (WORKING & PICTURE & (Ev == txt))
#define Cand_1 (WORKING & TEXT & (Ev == txt))
#define Cand_2 (WORKING & ON & (Ev == mute))
#define Cand_3 (WORKING & OFF & (Ev == sound))
#define Cand_4 (WAITING & STANDBY & (Ev == out))
#define Cand_5 (WAITING & DISCONNECTED & (Ev == in))
#define Cand_6 (WAITING & STANDBY & (Ev == on)) & \
               !(WAITING & STANDBY & (Ev == out))
#define Cand_7 (WORKING & (Ev == off))               & \
               !(WORKING & PICTURE & (Ev == txt)) & \
               !(WORKING & TEXT & (Ev == txt))    & \
               !(WORKING & ON & (Ev == mute))     & \
```

```
                !(WORKING & OFF & (Ev == sound))
#define Cand_8 (WORKING & (Ev == out))          & \
                !(WORKING & PICTURE & (Ev == txt)) & \
                !(WORKING & TEXT & (Ev == txt))    & \
                !(WORKING & ON & (Ev == mute))     & \
                !(WORKING & OFF & (Ev == sound))

proctype STEP()
{do
::
atomic{

if
:: Q_txt -> Ev = txt; Q_txt = 0
:: Q_mute -> Ev = mute; Q_mute = 0
:: Q_sound -> Ev = sound; Q_sound = 0
:: Q_out -> Ev = out; Q_out = 0
:: Q_in -> Ev = in; Q_in = 0
:: Q_on -> Ev = on; Q_on = 0
:: Q_off -> Ev = off; Q_off = 0
fi;

if
::Cand_6 -> WAITING = 0; STANDBY = 0; WORKING = 1; PICTURE = 1; ON = 1;
::Cand_7 -> WORKING = 0; PICTURE = 0; TEXT = 0; ON = 0; OFF = 0;
            WAITING = 1; STANDBY = 1;
::Cand_8 -> WORKING = 0; PICTURE = 0; TEXT = 0; ON = 0; OFF = 0;
            WAITING = 1; DISCONNECTED = 1;
::Cand_0 -> PICTURE = 0; TEXT = 1;
::Cand_1 -> TEXT = 0; PICTURE = 1;
::Cand_2 -> ON = 0; OFF = 1;
::Cand_3 -> OFF = 0; ON = 1;
::Cand_4 -> STANDBY = 0; DISCONNECTED = 1;
::Cand_5 -> DISCONNECTED = 0; STANDBY = 1;
::else -> skip
fi}
od}

init
{
atomic{
/* initial configuration */
WAITING = 1; STANDBY = 1; DISCONNECTED = 0; WORKING = 0; PICTURE = 0;
TEXT = 0; ON = 0; OFF = 0;
}
run STEP()
```

}

**Discussion**

- **Statecharts subset** The subset of UML statecharts considered covers the aspects related to concurrency and state hierarchy. Variables, history states, structured events and completion transitions are not covered but can be handled conceptually.

- **Multiple statecharts** The translation does not consider multiple statecharts with multiple input queues, one for each object, which communicate with each other. However, the authors claim that the translation can be easily adapted to handle multiple statecharts as well.

- **Complexity** The semantic rules and the translation scheme are more involved than the work in the previous section.

### 2.2.3 Translating UML statecharts to PROMELA/SPIN using vUML [Lilius et al, 1999]

The work by Lilius and Paltor [36] discuss a formalisation of UML state machines for translation to PROMELA/SPIN as part of their verification tool vUML [43]. The main features of this formalisation are listed below.

- The work has two parts:
    - A formalisation of the *structure* of UML state machines that is simple and declarative, and allows one to formulate the transition selection algorithm.
    - A formalisation of the *operational semantics* of UML state machines that is relatively complete, covering all its interesting aspects.

- The verification performed by the tool vUML is limited to the automatic verification performed by SPIN without user intervention, viz. deadlock checking and some robustness checks.

- The formalisation of the structure of UML statecharts uses linear ground terms over a signature with states as operations to describe state configurations. Transitions are defined as tuples of source and target configurations and a label, which defines the trigger, guard and the effect of the transition.

- The UML run-to-completion (RTC) step is modelled by an algorithm that calls the operations in an abstract data type (ADT) modelling the event queue.

**Discussion**

- **Statecharts subset** The subset of UML statecharts considered is bigger than in most other works.

**Scope of Verification** The verification performed in SPIN requires no user intervention, and is therefore limited – essentially deadlock detection.

Translation scheme The details of the translation to PROMELA are not provided in the cited papers.

## 2.3  The STATEMATE Verification Environment [Damm et al, 2000]

This work is reported in a series of papers by Werner Damm and his coworkers at OFFIS, University of Oldenburg [7, 8, 10], which describe the transition from a prototype verification system for STATEMATE to a commercial product from i-Logix. The salient features of the STATEMATE Verification Environment are listed below.

1. The environment uses an intermediate language called SMI for translating STATEMATE models into the input language of a model checker.

2. The current version (2000) is based on a tight integration with the VIS model checker [20] and the CUDD BDD package [50]. VIS is a BDD-based symbolic model checker that uses CTL to specify system properties.

3. The underlying verification technology is completely hidden from the user, using push-button analysis techniques and visual specification of properties in the form of a pre-defined specification pattern library. The pre-defined properties can be used to express both correctness properties of the design and assumptions about the environment.

4. To perform model analysis by symbolic model checking, the FSM describing the model's behaviour is extended automatically with *observers*, which allows the specification of robustness properties as atomic propositions $p$. These propositions are then checked using simple CTL formula $AGp$ for invariants. A counterexample path for this correctness formula can be used to drive the STATEMATE simulator.

5. The correctness properties specified in the user interface are translated automatically into temporal logic formulas, while assumptions about the environment are translated into observer automata.

6. Verification is done by adding the observers for the assumptions and fairness constraints to the model, and performing model checking using VIS.

7. The verification environment supports push-button analysis for verification of the following robustness properties of STATEMATE designs:

    - simultaneous activation of conflicting transitions
    - multiple write accesses to a single data item in the same step
    - parallel read- and write-access to the same object

8. Simple reachability mechanisms are provided to drive the simulation of a STATEMATE model to some user specified state or property.

9. To overcome the state explosion problem associated with model checking, the environment uses both *compositional reasoning* and *abstraction* techniques.

10. The verification environment computes the *cone-of-influence* (COI) of a model $M$ with respect to a property $\phi$, which restricts the model to only those variables that may influence the truth of $\phi$. The propositional abstraction supported in the STATEMATE verification environment provides a mechanism to automatically compute an over-approximation $M_\alpha$ with respect to a set of variables chosen by the user from within the COI. Both the COI and the computation required to to build $M_\alpha$ are performed on the intermediate representation of the model in SMI before translation to VIS. The abstraction technique can also handle models containing infinite objects.

11. The verification methodology is to use repeated COI reductions, abstractions and model checking. If the abstracted model satisfies the property, then so does the original model. If the model checking reports failure of the property, the counterexample is analysed, either to conclude that it is a genuine one or to identify further refinements in the abstraction, in case it is a spurious one.

## 2.4 Verification of Statecharts using Esterel Tools [Seshia et al 1999]

In [49], a method of translating statecharts to Esterel is proposed. A prototype implementation has also been developed by the authors. The aim of this translation is to extend the powerful verification and code generation tools of Esterel to statecharts. The important features of this proposal is:

1. The STATEMATE semantics of statecharts is used.

2. Since Esterel is deterministic, only the deterministic fragment of statecharts is considered for translation.

3. Almost all the features of statecharts, like history and inter-level transitions, are considered in the translation.

4. In the translation of a statechart, there is an Esterel module corresponding to each OR-state. The module simulates the behaviour of the state.

5. The signals of statecharts are translated to corresponding Esterel signals. In addition, each Esterel module has a set of signals corresponding to every transition in the state machines. The latter signals are used to implement the transfer of control from one module to another.

6. A special STEP signal is used for implementing the super step semantics of statecharts.

There are a number of problems with the proposed scheme:

- Since each OR-state is translated into a module, there is a state space explosion which would restrict the size of models that can be verified.

- Esterel tools have a very limited verification support and hence complex property-based verifications cannot be carried out using the tool.

- There is not much experience in using the tool in an industrial setting and hence it is not clear whether it will scale up.

- When the verification fails in the translated code, there is no traceability to the original model.

## 2.5 Verification of Communicating Reactive State machines [Ramesh et al]

In [46], a pictorial language based upon statecharts, called Communicating Reactive State Machines (CRSM), has been proposed for programming distributed real-time control applications. A comprehensive environment for developing and verifying CRSM descriptions has been developed [13, 53]. The environment enables editing and simulation of CRSM descriptions. Formal verification of the designs can also be carried out in this environment. Verification is based upon a translation of CRSM designs to PROMELA code. Properties are specified using distributed observer automata. Distributed observers are state machines, one for each CRSM node, which check whether any safety constraint is violated. The state machines of the distributed observers can communicate with each other and hence can specify both global and local constraints. The translation of CRSM into PROMELA code consists of the following steps:

1. Each node is translated into a PROMELA process.

2. The PROMELA process corresponding to a node (the node process) executes repeatedly at discrete intervals of time, the different transitions of the node.

3. At each step, the node process performs an atomic sequence of transitions that are triggered by the external inputs or the internal transitions from the same node.

4. The environment is modelled as a different process which controls the execution of the node process.

5. The observer process is also translated as part of the node process. A special kind of labels are generated in the PROMELA code which are used in the verification.

6. The properties for verification are automatically generated and fed to the SPIN model checker.

7. If the verification fails, traceability to the original CRSM designs is also built into.

Related to statecharts verification, there are a few problems with the tool:

- The semantic of CRSM is based upon Esterel rather than classical statecharts.

- Only safety properties can be verified using the tool.

- More experiments are needed to see the scalability of the tool for large industrial applications.

## 2.6 Exploiting Behavioural Hierarchy for Efficient Model Checking [Alur et al 2002]

In a series of papers [1–6] Rajeev Alur and his coworkers have proposed a method for model checking of hierarchical state machines by exploiting the modularity in the design. This is in contrast to the above approaches which translate statecharts specifications into the input languages of existing model checkers, thus losing the hierarchical structure in the input specification.

The input language to the model checker in this work is based on *hierarchic reactive modules* (HRMs) [4], a variant of statecharts where the notion of hierarchy in behaviour descriptions is *semantic* rather than *syntactic*. More precisely, HRMs have an observational trace-based semantics that allows defining a refinement preorder on hierarchic states. Further, HRMs support extended state machines where the communication is via shared variables. The central component of an HRM is a *mode*, which roughly resembles an OR-state in a statechart. A mode consists of local and global variables, well-defined control points, classified into entry and exit points, and submodes that are connected with each other by transitions. The transitions are labelled with guarded commands that access the variables according to natural scoping rules. The transitions can connect to a mode only at its entry/exit points, unlike in statecharts. Thus a mode is a black box whose internal structure is not visible from outside. The mode has a *default exit* point, whose outgoing transitions are applicable at all control points within the mode and its submodes. The default exit retains the history, and the state upon exit is automatically restored by transitions entering the default entry point. The operational semantics of modes specifies that transitions are executed repeatedly until there are no more enabled transitions. The transitions are tried for execution inside out, i.e., internal transitions have higher priority over the group transitions of the enclosing mode. The language allows multiple instantiations of a single mode, thus promoting reuse. The behaviour of a mode can be viewed as a game in which the environment transfers control to the mode at one of its entry points, and the mode transfers the control back to the environment at one of its exit points. These macro-transitions from an entry point to an exit point are used to associate a set of executions and a corresponding set of traces with a mode. The traces of a mode can be constructed from the traces of a submode, giving a denotational semantics of a mode.

The model checker, whether enumerative or symbolic, works by performing a reachability analysis of the input HRM model, as follows.

1. The model is parsed into an internal representation that directly reflects the hierarchical structure in the input.

2. The authors have implemented an enumerative checker based on depth-first search and a symbolic search that uses BDD packages from VIS [20].

3. The enumerative algorithm has the following features.

    (a) It takes as input a set of top-level modes and a set of global variables that these modes can read and modify. The property to be verified (say, an invariant) is also an input. Top level modes are assumed to be sequential and

run concurrently by interleaving their macro-steps. A state of the system consists of the values of all the global variables and the state of each of the top-level modes.

(b) In each round one of the modes may modify the variables and change its own internal state. The set of states of the system can be viewed as a directed graph where, if $s$ and $t$ are states of the system, $(s, t)$ is an edge in the graph iff $s$ yields $t$ after one round of execution.

(c) To search all states, a depth-first search is performed on the graph beginning from an initial state. For each state encountered in the search the desired invariant is checked. If the invariant doesn't hold then the DFS algorithm supplies a path in the graph from an initial state to the state violating the invariant.

(d) The set of states visited is stored in a hash table, so that membership can be decided in constant time.

(e) The set of successors of a state is computed by examining the modes. The hierarchical structure of the modes is retained throughout the search. The analysis algorithms attempt to exploit this structure in different ways:

   i. The transition relation is maintained indexed by the modes and their control points for quick access to potentially enabled transitions.
   
   ii. States are represented as stacks of vectors rather than vectors. This is useful for handling priorities of transitions.
   
   iii. Variable scoping is also exploited for efficient representation of state space. The stack representation leads to smaller mode state sizes compared to a system where all variables are global. The total size of the state of a mode is proportional to the depth of its hierarchy.
   
   iv. The stack structure of a mode's state also allows conserving memory by sharing parts of the state.
   
   v. When one mode is instantiated in many places in an HRM all of them exhibit the same behaviour when their global variables are the same. The model checker avoids recomputing a mode's behaviour if another instance of that mode has already been searched for an equivalent context.

4. The symbolic search algorithm has the following features.

(a) The algorithm does not assume that the top-level modes are sequential. A state in this case is not a stack, but a map (context) from variables to their values. This context varies dynamically, depending on the currently accessible variables.

(b) Instead of flattening the input HRM mode into a transition relation and then represent the reached states and the transition relation by ordered multi-valued decision diagrams (MDDs), the algorithm keeps the MDDs in a decomposed way, as suggested by the modular structure. This results in a more efficient use of memory.

(c) The reachable state-sets are represented not by a single MDD, but as a mapping of the currently reached control points to their associated reached region MDD. Such a representation allows us to partition the state space intuitively, with each region containing all the states with the same control point.

(d) The transition relation is represented as a map from control points to a list of pairs containing destinations of edges along with MDDs encoding the corresponding guarded commands. In this representation each transition is much smaller than the counterpart used in a flat representation. This is possible because local variables in distinct modes are not simultaneously active and the mapping provides information on control points. The typing and scoping information of the original model is maintained during compilation to achieve this economy.

(e) The reachability computation computes reachable states at each control point. When a top mode $M_i$ gets control for the first time, it starts the image computation from its entry point by following the transitions until the control gets stuck.

(f) The image computation returns an MDD $S_i$ that contains the information about where and how the control inside $M_i$ gets stuck.

(g) In the first iteration, each top mode is given a chance to do the first image computation. The next iteration is started by building a current onion ring for top mode $M_i$ based on the stuck sets $\bigcup_i S_i$. The current onion ring is a map from the control points where the control became stuck during the last image computation at $M_i$ to newly reached states obtained from image computations at top modes other than $M_i$.

(h) By applying the image computation to the current onion ring at $M_i$, the control may continue from those stuck control points.

(i) The algorithm terminates if all the onion rings for top modes are empty i.e., no new states can be reached at any control point.

**Discussion**

1. Unlike statecharts, HRMs have well-defined entry and exit points and the transitions can connect to a mode only at its entry/exit points. A mode in an HRM, unlike a state, can be instantiated more than once, leading to sharing. The communication between the top-level modes is via shared variables. The local and global variables in HRMs follow the usual scoping rules in block structured languages. Actions (guarded commands) are associated with only transitions and not modes or control points. All this makes the model substantially different from the statecharts model, and difficult to apply the technique to traditional statecharts.

2. The authors state in [1] that the experimental evidence to support the thesis that the proposed solution leads to more tractable analysis than compilation into a non-hierarchical checker is small. However, they claim that there is adequate conceptual evidence in their favour.

## 3 Future Directions

We have reviewed a number of approaches for verification of statecharts-like descriptions. All the tools except [7] are prototypical academic tools accepting a subset of features and having a wide variety of semantics ranging from STATEMATE semantics, UML semantics to synchronous semantics. All the tools translate statechart descriptions into another modelling language which can then be verified. The back-end modelling languages, except the one in [3], do not exhibit the high level features like hierarchy and synchronous or step semantics of statecharts. In the process of translation, these high level features are removed. For example, hierarchical statecharts are flattened leading to large state space. This will come in the way of scalability of the proposed approaches. On the other hand, the input language of HRMs in Alur et al, though retaining the hierarchical structure, is significantly different from statecharts. Also, except for the work reported in [7, 41, 46], there is no traceability between the front-end and back-end design models, which is an important requirement, if one wants to debug the design based upon the verification report. Finally, very few of the tools, with the exception of the STATEMATE Verification Environment described in Secrefsve, have been tested on actual industrial examples.

Our future work will focus on all these aspects to arrive at a more mature tool for verification of statecharts. Here we identify some research issues that would address the gaps in the current approaches to verification of statecharts.

**Verification of statecharts without flattening the hierarchy** Although the work of Alur et al [1–6] cited above is a step in this direction, the hierarchical reactive modules (HRMs) model is significantly different from statecharts. The challenge would be to adopt the ideas presented to reflect the semantics of statecharts. Statecharts do not have the block structured push-down hierarchy displayed by HRMs. The hierarchy is broken by interlevel transitions and higher priority for outer transitions in the STATEMATE semantics (the UML priority scheme is very much like HRMs). Moreover, since all variables in statecharts are global, there will be no saving in state space size by having a stack of vectors as a representation of the state space. The enumerative search algorithm proposed in [1, 3] will have to be suitably modified to take into account these differences.

**Compositional Verification based Approaches** In a recent paper [45], some efficient techniques for verification of synchronous programs expressed in the language Argos [38] have been proposed. In this work, the conditions under which the results of verification of subcomponents can be extended to the verification of the whole system or program are explored. Here the subcomponent could be a concurrent automaton or top-level automaton with some of the hierarchical states being converted to basic states (i.e., the inner state machines removed). Verification of the latter is easier than the entire state machine, as the number of concurrent or hierarchical components contributing to the global states is reduced. Obviously, not all properties can be proved using this approach. If the input formula satisfies certain conditions, like it is local (referring to a fewer localised events) and it does not state any negative properties about events and states not present in the subcomponents, then the result holds.

Based upon this idea, a compositional verification strategy is proposed: given a property to be verified, break the property into sub-properties and verify the individual properties separately. Verification of the latter is quite likely to be restricted to relatively few components and hence would be less expensive than verifying the whole property on the whole system.

Currently, the method works for Argos [38]. The Argos semantics is a bit different from statecharts and there is a need for extending the above methods to statecharts.

**Static Analysis, Slicing and Abstraction based Approaches** Another way of reducing complexity of verification is as follows: based on the property to be verified, the input statechart can be reduced by using slicing or cone-of-influence reduction. This is a promising approach which has not been sufficiently investigated, although there are some early work [7, 19]. Abstraction techniques are a powerful mechanism for reducing infinite or large state spaces to small ones suitable for model checking. But abstraction is not an automated process and requires discharging proof obligations. This requires an interaction between theorem proving and model checking which is an active area of research.

**Assumptions about the environment** Assumptions about the environment of a statechart can lead to substantial reductions in state space size. These assumptions could be on the sequence of inputs from the environment, as well as the size of the data.

**Refinement** Another possible approach to verification would be to refine a high level abstract specification into a detailed design specification. Both these specifications would be in the form of statecharts. Verification would amount to checking that the design implements the abstract specification by specifying refinement maps between the two.

# 4 Acknowledgements

The authors thank Sushil Birla, Tom Fuhrman and Shengbing Jiang of General Motors and Mathai Joseph of TRDDC, Pune for many discussions on this subject. The second author thanks the Centre for Formal Design and Verification of Software for the support.